# Analysing the anisotropy in morphological evolution and readjustment effects in cluster-cluster aggregation of AuNPs using Shannon entropy

Anurag Singh, Anushree Roy and Amar Nath Gupta*

Biophysics and Soft Matter Laboratory, Department of Physics, IIT Kharagpur, Kharagpur-721302, India

*Corresponding author's email id: ang@phy.iitkgp.ernet.in

## Abstract

We have used information theory analogue of entropy, Shannon entropy, for estimating the variations during the isotropic and anisotropic AuNP fractal growth process. We have firstly applied the Shannon entropy on the simulated fractal aggregates obtained from DLA model with noise reduction scheme. In conventional noise reduction scheme used in past, the growth process of identical particles was performed and no effect of the evolving cluster on the incoming particle was considered, hence the noise is reduced in discrete amount and do not account for the noise fluctuations present during the morphological evolution of the fractals. Experimentally, the distribution of particles aspect ratio is seen to be Gaussian in nature; hence in order to account for the particle aspect size ratio originated noise fluctuations present during the growth process, the noise reduction parameter $m$ was taken from Gaussian distribution. The standard deviation of the Gaussian distribution accounts for the amplitude of noise fluctuations present in the system. The Shannon entropy is observed to capture these fluctuations of noise in the system. The tool was further applied on experimental images for morphological evolution of both regular and anisotropic AuNPs fractals. The Shannon entropy is shown to capture the emergence of the anisotropic morphological evolution. The imaging tool was further found to be promising for capturing the readjustment effects during cluster-cluster aggregation. These analysis shows that Shannon entropy can be used for capturing entropic changes during different aggregation processes.

**Introduction**

Nanoparticles, particularly gold nanoparticles (AuNP), show unusual physical (easy surface modification, electronic, magnetic and optical) and chemical (stability, catalytic) properties. The characteristics of self-similar fractal structure of aggregated gold colloids have been studied for decades [1-3]. Due to large surface to volume ratio of these colloidal particles the resonance of collective oscillation of electron density at the surface under certain frequencies of visible radiation exhibits a dominant effect. This surface plasmon resonance of these particles marks these systems interesting from physics point of view and also makes them potential candidates for various applications. The light induced self-assembly of AuNP has shown various applications in recent past, self-erasing ink [4, 5], micrometer sized supercrystals[6], photoswitches [7, 8], aggregation based immunoassay[9], biomarker[10-13], and sensors[14, 15], in therapy via photothermal, photodynamic, therapeutic agents, drug carriers and toxicity therapy, to name a few.

The self-organization and aggregation of these nanoparticles is driven by the minimization of the free energy of the system mostly through diffusion and other physical conditions. The dynamics of self-assemblies of molecules have been described by many theoretical models, DLA [16], RLA[17], RSA[18], percolation[19], Ising model[20] etc. Particles often possess geometrical symmetry or surface isotropy, leading to symmetric interactions among them that result in highly-symmetric clusters[21] and periodic lattices[22]. But, sometime these self-assembly turn out to be directional by breaking the surface symmetry. The directionality in the morphology depends on the noise fluctuations present in the system, which is, apart from thermal fluctuations, mainly controlled by the aspect ratio of the particles present in the system. Moreover for the synthesis of nanoparticle based drug delivery system, the aspect ratio of the particles present in the system plays an important role in deciding the size and shape of the drug carriers. In the first part of this study, we have addressed the contribution of the particles aspect ratio related noise in deciding the final morphology of the formed fractals.

The directional self-assembly has applications in optics, magnetics, and diagnostics. One of the examples is fractal antenna[23] which is approximately 20% more efficient than normal antennas, a self-similar design to maximize the length that can receive or transmit the EM signals within a given total surface area. As, both growth and the anisotropic fractals have equal importance from application point of view, it is important to capture the evolution of their growth processes. Moreover, in nature we have abundant of examples where both isotropic and anisotropic fractals can be observed which makes the morphological evolution of the study very important from the point of view of understanding the physical basis of existence of such morphologies. Here, we have tested the Shannon entropy for capturing the isotropic and the anisotropic growth process. We show that the Shannon entropy is capable enough to capture emergence of the directionality in the growth process.

The DLA[16] model is a particle-cluster aggregation model where a particle stick one after another onto a single growing aggregate. The model has been extensively used for describing various non-equilibrium field-induced growth phenomena such as pattern formation [24], viscous fingering[25] and electrodeposition [26]. However, the aggregates obtained using the Witten-Sander DLA model are more compact in comparison to the structures obtained from standard aggregation experiments. To account for this discrepancy, an alternative model of cluster-cluster aggregation, known as Diffusion-limited cluster aggregation (DLCA), was introduced in past, the algorithm of which is very similar to that of the cluster-particle DLA model [27].

In simulations, the cluster-cluster aggregation process starts with randomly placed individual particles in a box. The particles move randomly inside the box considering the periodic boundary conditions at the edges of the box. The particles stick irreversibly when they come in contact, which leads to the formation of dimer. Then the dimers can stick and move to other dimers or single particles. The small clusters moves randomly having velocity inversely proportional to its number of particles. The distribution of mean cluster size and the shape of the cluster, which evolve with time (shift from cluster-cluster aggregation to particle-cluster aggregation), can be described by Smoluchowski equation [28]. The process is continued until a large aggregate is formed inside the box. The aggregates obtained using simulations are quite different from the one obtained in real experiments. This is mainly because of the approximation in cluster-cluster aggregation model according to which the clusters stay rigid during their diffusive motion and do not rearrange after sticking. However, experimentally, the readjustments effects have been observed in many experiments where smaller clusters have been observed to reorient around the bigger cluster in search of the proper orientation to stick in. We have used Shannon entropy to capture the reorientation effects during the cluster-cluster aggregation in AuNPs. The Shannon entropy has proved to be promising imaging tool for capturing the readjusting effects.

**Methods**

**1. Simulation**

We performed the simulation on the square grid of size 500X500 grid points with 4000 particles. In the simulation, our approach is very similar to the DLA simulations which had been performed in the past for obtaining the snowflake-like structure[16]. In DLA simulations, the particle stick to the nucleation point in the first attempt itself. In such simulations, a nucleation point is placed in the centre of the square lattice and other particles are generated at some other lattice sites. The generated particle is allowed to perform the random walk on the square lattice. If the particle comes near to the nucleation point, the index of the lattice point is increased to 1 and the particle is said to be attached to the nucleation point. To observe the structural transition from regular isotropic to anisotropic growth, hence to study the morphological evolution of the fractals in both the growth cases, the index of the neighbouring unoccupied lattice was increased to $m$ rather than 1 for the particle to get attached to the growing structure. In our simulation, the isotropic growth and the anisotropic growth were obtained with the noise reduction parameter of $m=1$ and $m=8$ respectively.

The complete transition from snow-flake like structure to anisotropic fractals can be observed by increasing the value of the noise reduction parameter in positive integers starting from 1, where $m=1$ corresponds to standard DLA process. In our case, the sufficient level of anisotropy required for the study, was observed at $m=8$, hence the $m$ was varied from 1 to 8. The sufficient level of anisotropy was decided by the similarity in the morphologies of the simulated fractals for $m=8$ and the anisotropic fractals observed experimentally. To incorporate the Gaussian noise fluctuations in the system, the $m$ values for every next particle in the growth process was drawn from the Gaussian distribution with the starting nucleation centre having the value given by the mean of possible noise reduction parameter values. The amplitude of the noise fluctuations in the system was decided by the standard deviation of the Gaussian distribution from which the $m$ values were drawn.

## 2. Experiments

The gold colloid was prepared using the well-established citrate reduction method mentioned elsewhere[29]. In short, 1.4 ml of 100 mM solution of $HAuCl_4$ was dissolved in 98.6 ml of water. The mixed solution was heated till it boiled. 10 ml of sodium citrate solution of 1% (1 mg in 100 ml) was added to the above solution as a reducing agent, with continuous heating and vigorous staring for 30 minutes. The colloidal sol was wine red in color. The surface plasmon resonance (SPR) measurements of the colloidal sols were carried out using the spectrometer Evolution-210 (Thermofisher, USA). Approximately 2 μl of the colloidal solutions were dropped on a thoroughly cleaned glass plate then the drops were allowed to dry under ambient conditions. An optical microscope (Nikon EPLIPSE Ti) in transmission mode with 100X objective lens was used to record the isotropic and anisotropic morphological evolution of the fractals in video form with speed of 100 frames per second. For capturing the cluster-cluster aggregation process, an optical microscope (BX41, Olympus, Japan) with 10X objective lens was used.

### a. Characterization of Gold Colloidal Particles

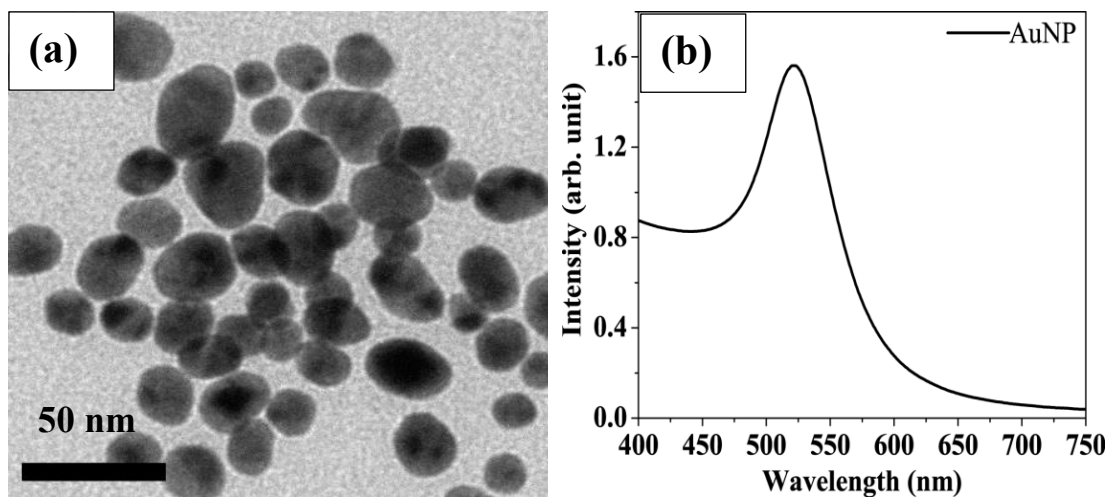

**Figure.1** (a) Characteristic TEM images of gold nanoparticles. (b) Variation of SPR band in the sol.

The characteristic TEM image of gold particles in as-prepared is shown in Fig. 1. We have analyzed 40 particles for each sample. We find that particles are nearly spherical with average aspect ratio of 0.87±0.09 in the as-prepared gold colloid.

The colloidal sol was further characterized by measuring the SPR band of the particles [30, 31]. The maximum of the SPR band of as prepared sol appears at 523 nm. As the particles are nearly spherical and we observe a single absorption peak over the entire visible range of radiation (note that two peaks related to longitudinal and transverse modes are expected to be observed for the ellipsoidal particles), under electrostatic approximation the absorption coefficient of a small particle under electromagnetic radiation is given by $C_{abs} = 12x \dfrac{\varepsilon'' \varepsilon_m}{(\varepsilon' + 2\varepsilon_m + \frac{12}{5}x^2\varepsilon_m)^2 + \varepsilon''^2}$ , where $\varepsilon'$ and $\varepsilon''$ are real and imaginary parts of dielectric constant of gold, $\varepsilon_m$ is dielectric constant of the medium. $x = 2\pi a/\lambda$ with $a$ as radius of the particle and $\lambda$ as the wavelength of the incident radiation. $C_{abs}$ vary with $\lambda$ and exhibit a maximum. The maximum of the absorption peak depends on wavelength dependent values of $\varepsilon$, $\varepsilon''$ and the values of $\varepsilon_m$ and $a$. The value of $\varepsilon_m$ for water is 1.77.

**b. Fractal dimension calculation**

The fractal dimension of the fractals corresponding to the integer and fractional $m$ values were calculated using power spectra method. This method is the application of Fourier power spectrum method. The fast Fourier transform is used to Fourier transform the real space and the power spectrum $S(k)$ is related to the wavenumber $k$ as[32];

$$S(k) \propto k^{-\beta}$$

where $\beta$ is the slope of the log-log plot of the power spectrum and the wavenumber $k$. The fractal dimensions of the 2-D images were calculated using the relation, $FD = (8 + \beta)/2$, between the fractal dimension, $FD$, and the slope.

**c. Shannon entropy calculation**

Shannon introduced the concept of entropy in 1948[33]. It measures the uncertainty of a random variable in information theory. The image Shannon entropy quantifies the information content of the image. Mathematically, it can be given as;

$$S = -\sum_{j=0}^{2^{\beta}-1} p(a_j)\log(p(a_j))$$

Where $S$ is the Shannon entropy, bits/pixel; $\beta$ is the pixel depth of the image; $p(a_j)$ is the normalized probability of the occurrence of each gray level, which can be obtained by the histogram of the image. We calculated the Shannon entropy of both the simulated fractals and the experimental images. The "histim( )" Matlab routine was used for obtaining the histogram plot of the image which gave the normalized probability of the occurrence of each gray level, $p(a_j)$; "entropy( )" Matlab routine was used for calculating the Shannon entropy.

**Results and discussions**

**1. Gaussian noise fluctuations in fractal growth process**

Fig.2 shows the obtained fractals on the square lattice for noise reduction parameter values $m = 1, 2……8$ and corresponding fractal dimensions, calculated using power spectrum method, are mentioned in the Fig. 2 for each $m$ respectively. It is observed from Fig. 2 (a)-(h) that there is a transition from snowflake-like structure to the directional fractals with the increase in the value of $m$. The calculated fractal dimensions for directional fractals (see Fig. 2 (d)-(h)) were in agreement with on-lattice DLA fractal dimensions obtained in previous study[34]. Here, the number of particles in each cluster is same and no other influencing parameters have been employed for generating these fractal structures as was done in past[35, 36]. Even though the fractals obtained through this simple approach can give us the realisation of the anisotropy in the growth process, this does not account the possible noise fluctuations present in the system.

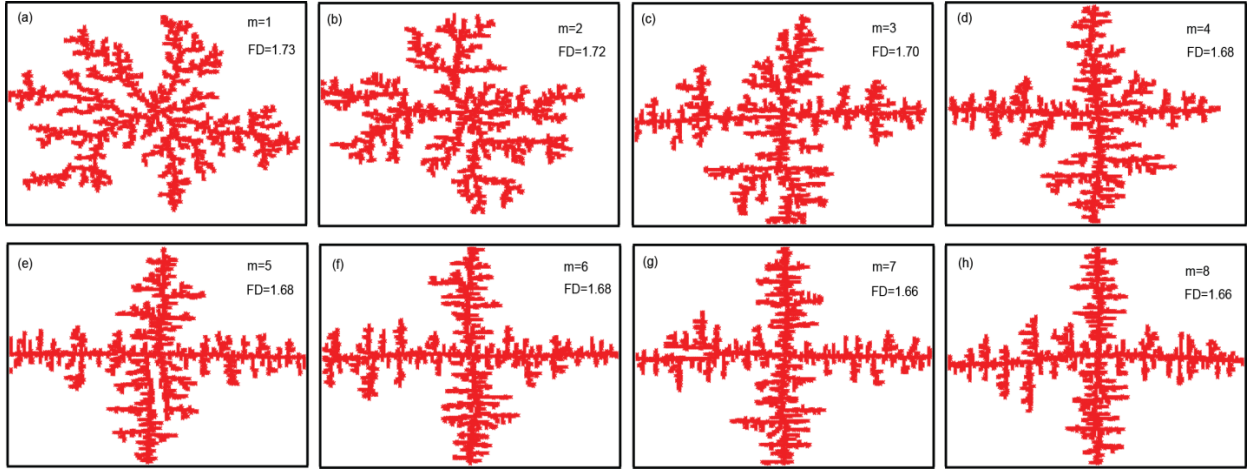

**Figure2 (colour online).** The figure shows the fractal structures obtained on the square lattice for $m$ =1, 2....8 respectively. The simulation was performed on 500X500 square lattices with 4000 particles. The fractal dimension of the fractal aggregates is also mentioned in the figures. It is evident that higher the value of $m$, greater is the directionality in the obtained fractal structure.

One of the noise contributions in the growth process is the noise mainly coming due to different shape and size of the incoming particle and the evolving cluster at every growth step. The growth process with particular value of noise reduction parameter depicts aggregation of identical particles with constant effect of the evolving cluster on the incoming particle. The increase in the noise reduction parameter results in the Monte-Carlo ensemble averaging of the site, which results in the reduction of noise and hence the anisotropic morphology in the on-lattice simulations. In more realistic case, the number of times a site should be sampled before a particle can finally attach depends on the evolved cluster to which the new particle is going to attach and the shape and size of the incoming particle. Hence, in reality every new particle is supposed to have different value of noise reduction parameter. We have incorporated this noise in on-lattice DLA model, by assigning mean value of noise reduction parameter ( taken from the range of possible values of $m$) to the nucleation point and a random $m$ value drawn from the Gaussian distribution, G ($\mu$=$m$, $\sigma$), to every new particle in the growth process. The choice of the Gaussian distribution is motivated from the fact that, as represented in the characterization section, the aspect ratio of the particles is following Gaussian distribution. Since the sampling of the space near the growing cluster should depend on the shape and size of the growing cluster, the $m$ values, which essentially represents that how many times the site should be sampled, should be drawn from the Gaussian distribution. In the conformal mapping model given by Hastings and

Levitov [37, 38], which represents the off-lattice simulation model of DLA, this is incorporated by the acceptance angle range between which the new particle can come and attach.

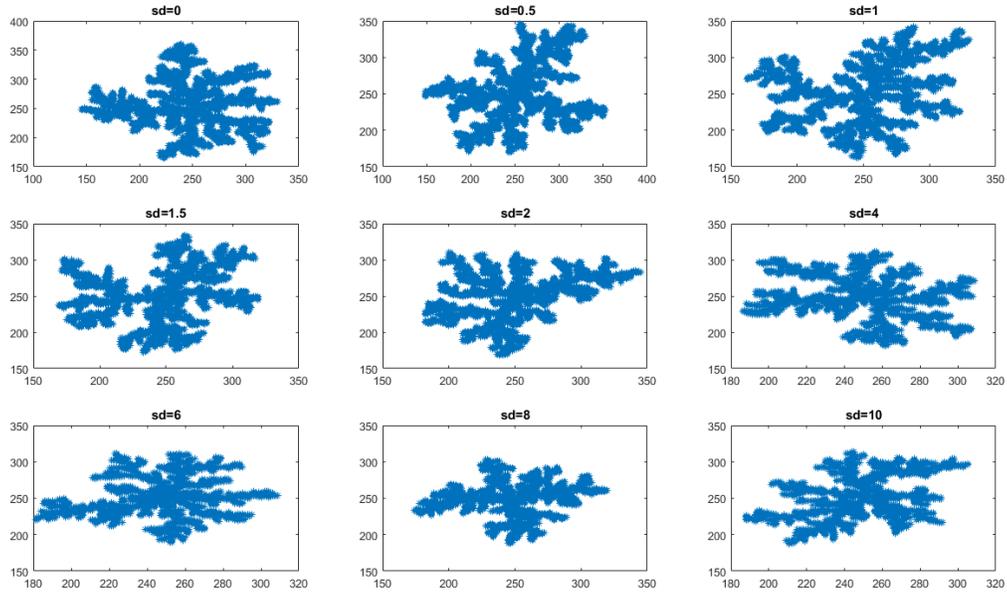

**Figure3 (colour online). F**ractal structures obtained on the square lattice for *m*=1 drawn from the Gaussian distribution with σ=0, 0.5, 1, 1.5, 2, 4, 6, 8 and 10.

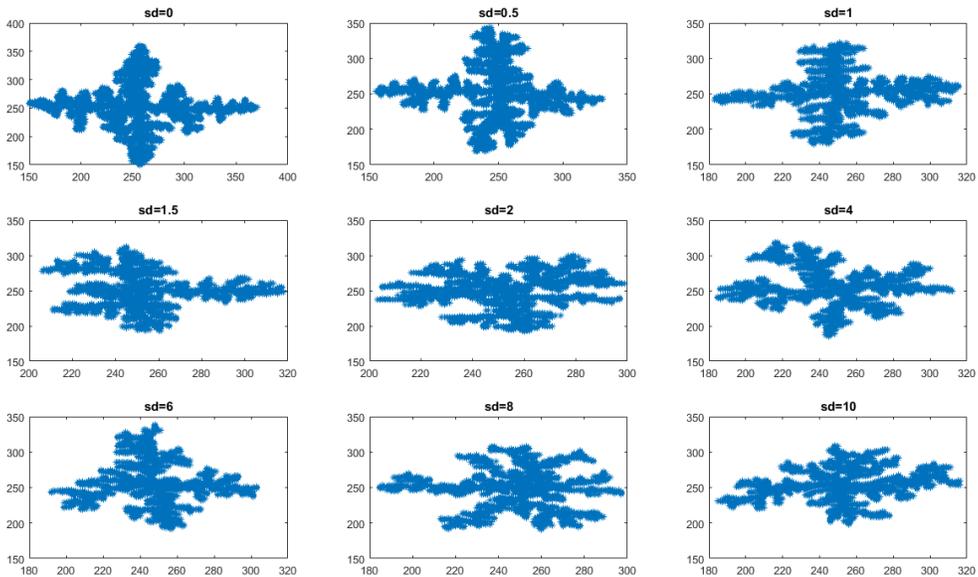

**Figure4 (colour online). F**ractal structures obtained on the square lattice for *m*=4 drawn from the Gaussian distribution with σ=0, 0.5, 1, 1.5, 2, 4, 6, 8 and 10.

Fig. 3 represents the fractals obtained with varying noise fluctuations of 0, 0.5, 1, 1.5, 2, 2, 4, 6, 8 and 10 for *m*=1. As for *m*=1 standard DLA process is expected, for all the noise fluctuations snow-flake like morphologies are seen. However, with increase in the noise level, more noisy growth is expected as is evident from the morphologies shown in Fig 3.

Fig. 4 represents the fractals obtained with varying noise fluctuations of 0, 0.5, 1, 1.5, 2, 2, 4, 6, 8 and 10 for *m*=4. For *m*=4 transition in the morphologies from isotropic to anisotropic is observed (Fig 2).With increase in the noise in the system, the transition gets delayed and for standard deviation value of 10, even for *m*=4, non-directional fractals is observed. It is noted that the directionality features starts fading and gets hidden in the noise after the noise fluctuation of 1.5.

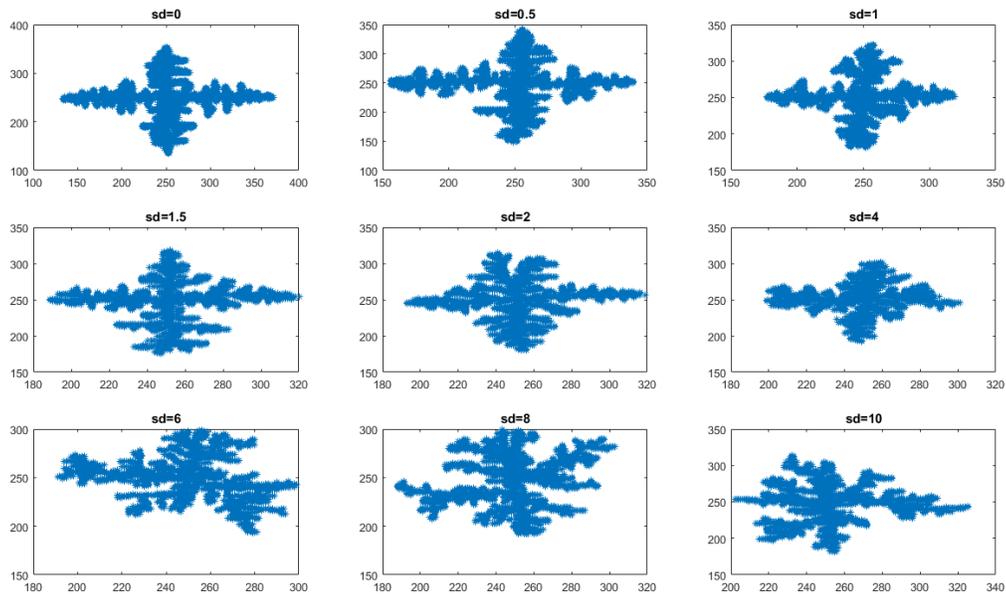

**Figure5 (colour online).** The figure shows the fractal structures obtained on the square lattice for *m*=8 obtained from the Gaussian distribution with σ=0, 0.5, 1, 1.5, 2, 4, 6, 8 and 10.

Fig. 5 represents the fractals obtained with varying noise fluctuations of 0, 0.5, 1, 1.5, 2, 2, 4, 6, 8 and 10 for *m*=8. For *m*=8 the directional morphologies are expected. With increase in the noise in the system, the directionality gets vanished and for standard deviation value of 10, even for *m*=8, non-directional fractals is observed. It is noted that the directionality features starts fading after the noise fluctuation of 6.

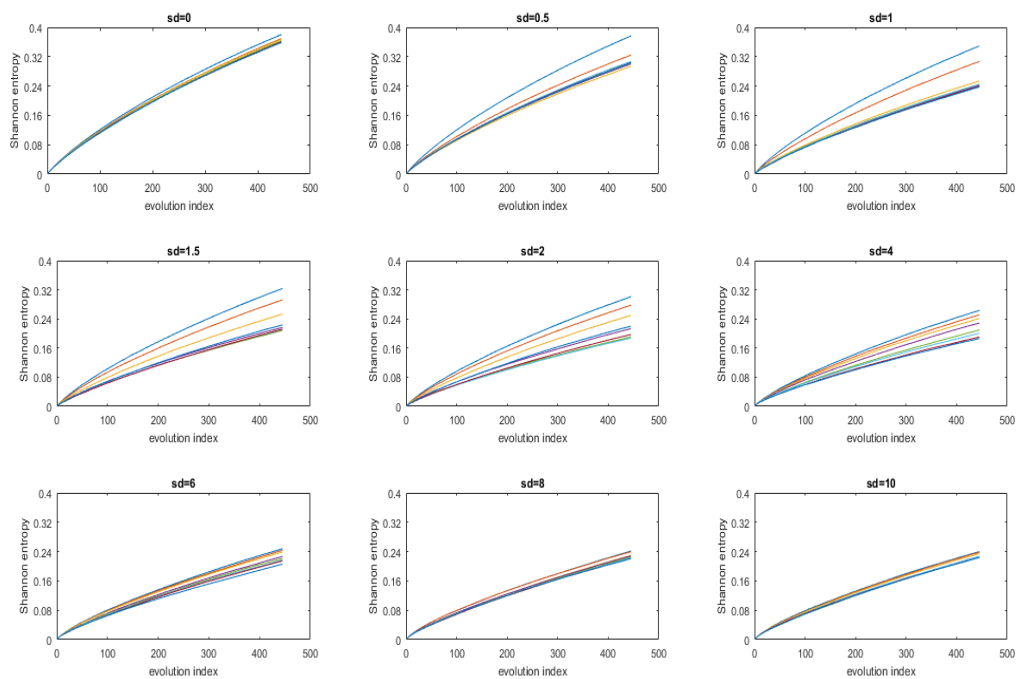

**Figure6 (colour online).** The variation in the Shannon entropy values for structures obtained on the square lattice for *m*=1, 2, ……8 represented by different color curves) drawn from the Gaussian distribution with σ=0, 0.5, 1, 1.5, 2, 4, 6, 8 and 10.

Fig 6 represents the Shannon entropy values of the snapshots of the images taken from the video capturing the morphological evolution for standard deviation of 0, 0.5, 1, 1.5, 2, 4, 6, 8 and 10 for all the values of *m*=1,2….8. With increase in the noise fluctuations, the separation between the curves for consecutive *m* values is becoming more in comparison to both when there is no noise in the system (sd=0) and when there is maximum noise in the system (sd=10). Moreover, the value of the Shannon entropy for case with maximum entropy has decreased by 40% for the case in comparison to that of the case for minimum noise (sd=0) in the system. The Shannon entropy shows similar curve behaviour as that of radius of gyration of the evolving cluster. However, with increase in the noise reduction parameter, the radius of gyration of the evolving cluster will increase in contrast to the observed decrement of the Shannon entropy with increase in the value of noise reduction parameter. This indicates that the fractals are becoming more stable with increase in the noise reduction parameter. This was further confirmed by calculating the interaction energy of the formed fractals with different noise reduction parameter values (Fig 16 (a)).

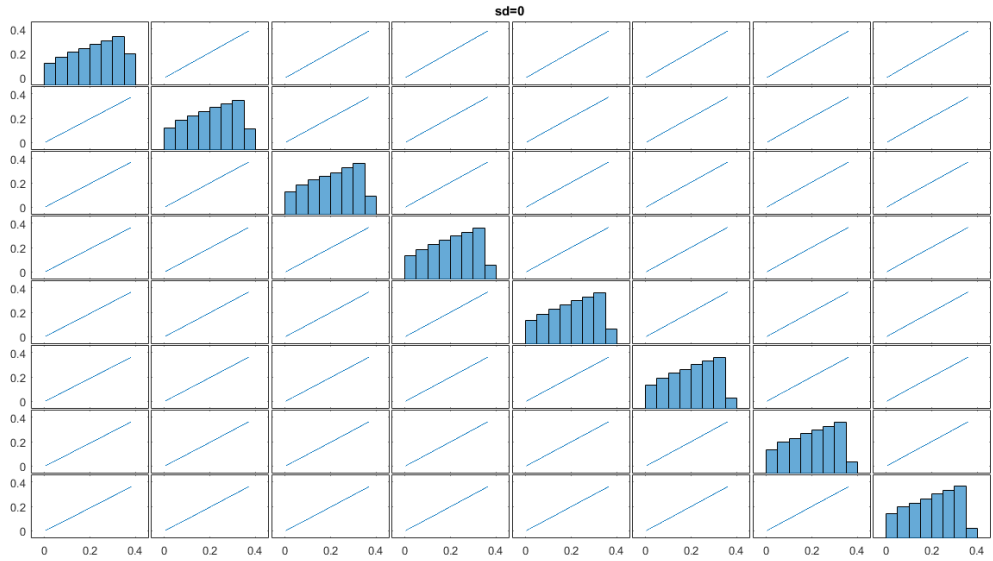

**Figure7 (colour online).** The *m*-matrix for σ=0 represents the interdependency between *m*=1, 2, ……8 drawn from the Gaussian distribution with σ=0. This represents the growth with identical particles, having same particle size and shape.

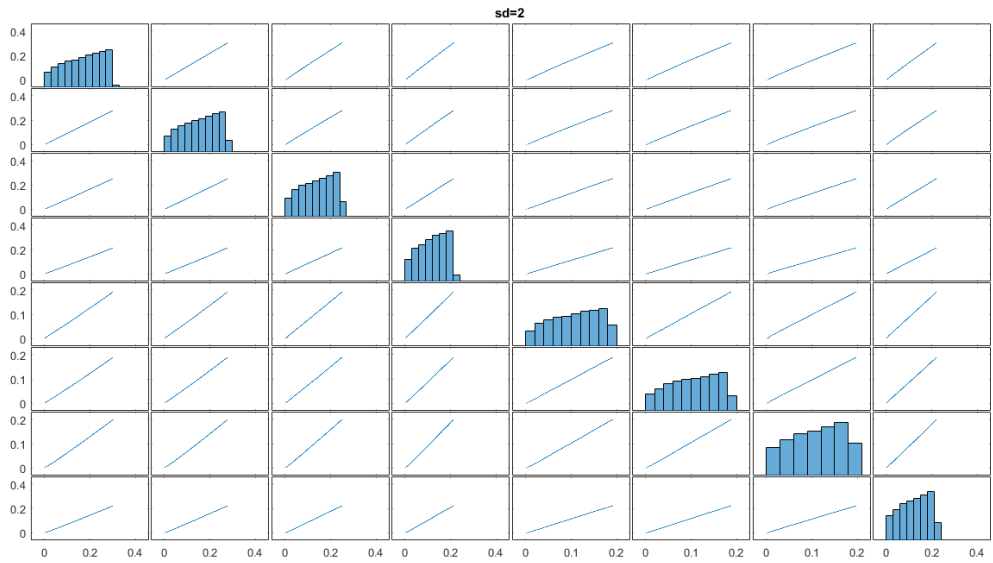

**Figure8 (colour online).** The *m*-matrix for σ=2 represents the interdependency between *m*=1, 2, ……8 drawn from the Gaussian distribution with σ=2. This represents the growth with particles having standard deviation of 2 in their aspect ratio.

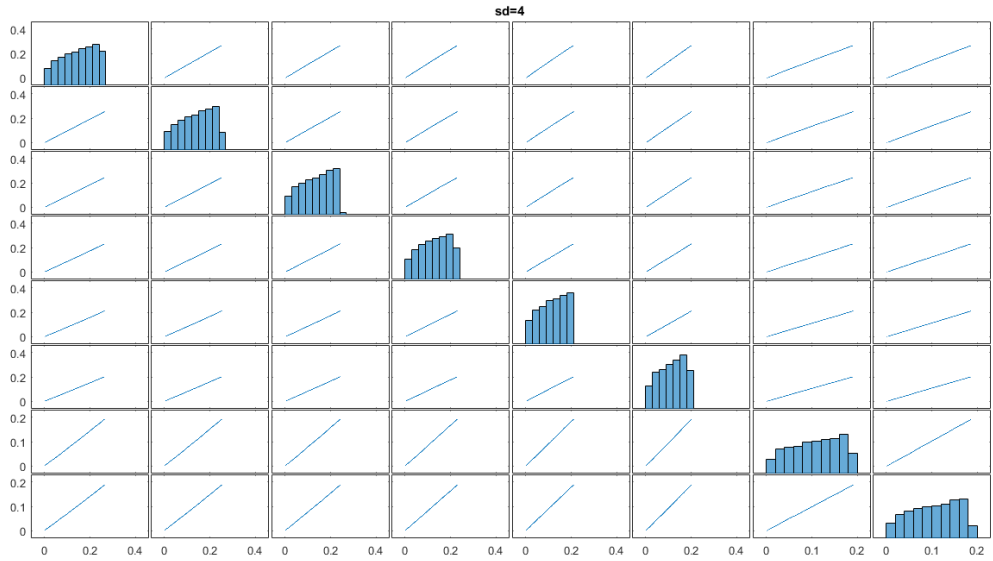

**Figure9 (colour online).** The *m*-matrix for σ=4 represents the interdependency between *m*=1, 2, ……8 drawn from the Gaussian distribution with σ=4. This represents the growth with particles having standard deviation of 4 in their aspect ratio.

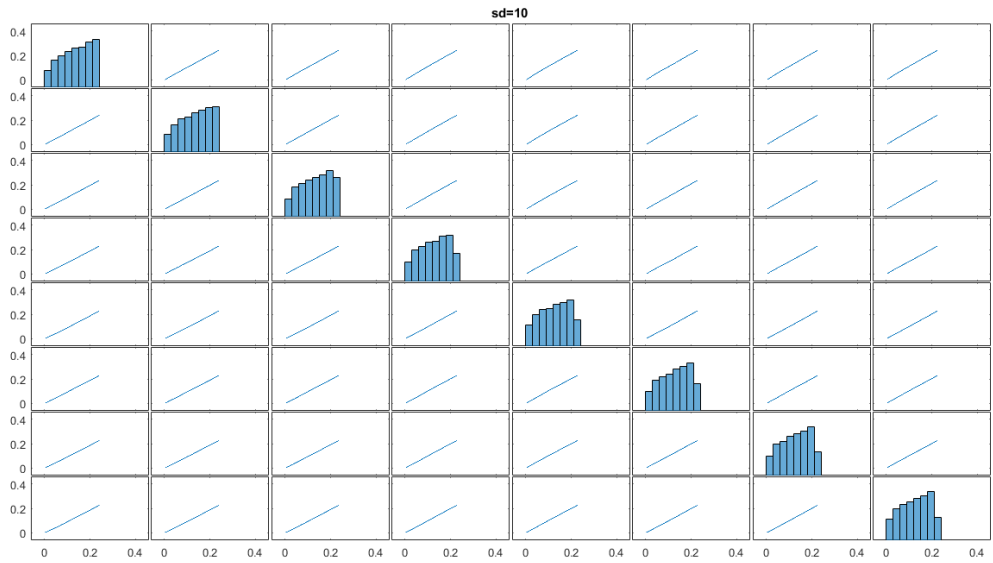

**Figure10 (colour online).** The *m*-matrix for σ=10 represents the interdependency between *m*=1, 2, ……8 drawn from the Gaussian distribution with σ=10. This represents the growth with particles having standard deviation of 10 in their aspect ratio.

Fig 8, 9 and 10 represents interdependency in the Shannon entropy values of the evolving fractals for standard deviation of sd =0, 2, 4 and 10. The diagonal elements represent the histogram plot of the Shannon entropy values obtained for that particular value of noise reduction parameter *m*. The off-diagonal elements depict the one-to-one correlation between the Shannon entropy values between the respective *m* values in the *m* matrix. It is observed that for no noise fluctuation (σ=0) and maximum noise fluctuation (σ=10) the slope of the straight lines showing the interdependency is almost same. But for the intermediate noise fluctuation value the interdependency showing straight lines have different slopes. This shows that in presence of noise fluctuations, the obtained morphologies can be different than that expected from the standard DLA process. Also, this represents that in real system there is a noise level after which no further directionality in the morphologies is expected. Moreover with increase in the noise fluctuation, the transition from non-directional fractals to directional fractals is delayed decided by the noise in the system.

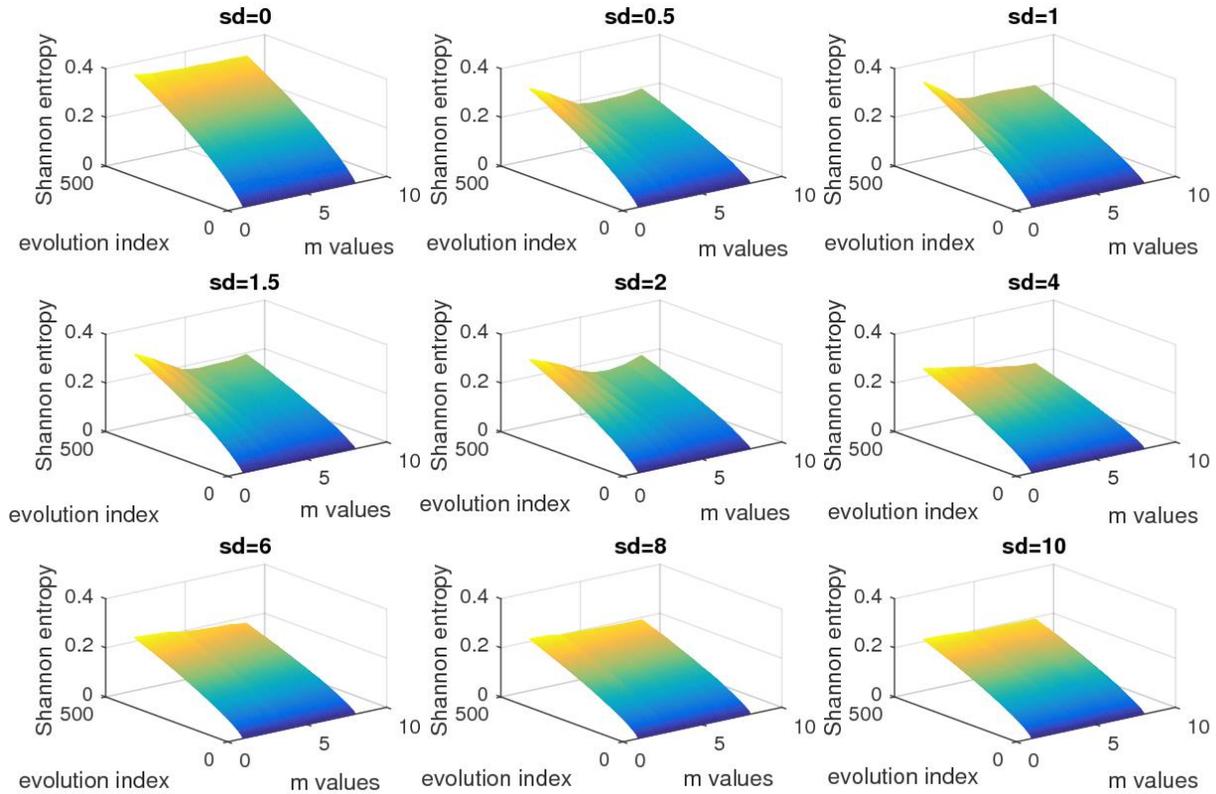

**Figure11 (colour online).** The 3D contour plot of the variation in the Shannon entropy values for structures obtained on the square lattice for *m*=1, 2, ……8 drawn from the Gaussian distribution with σ=0, 0.5, 1, 1.5, 2, 4, 6, 8 and 10.

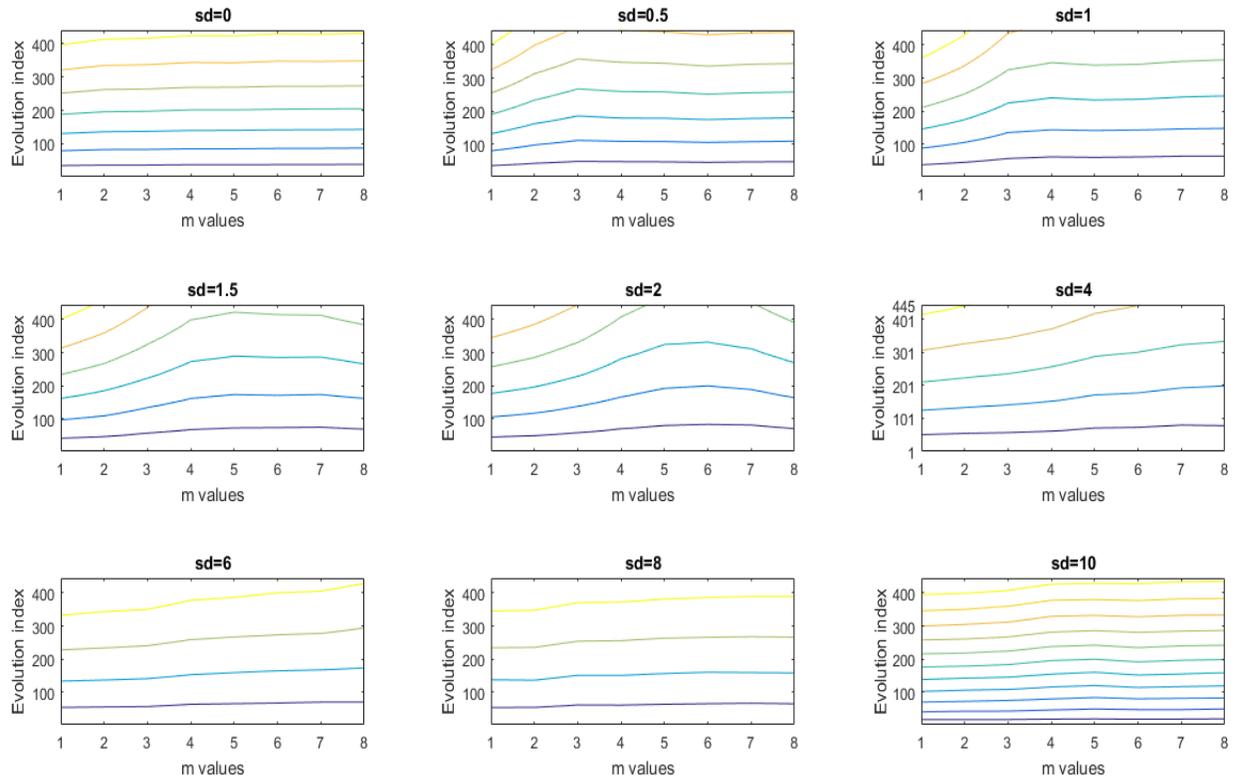

**Figure12 (colour online).** The 2D contour plot of the variation in the Shannon entropy values for structures obtained on the square lattice for *m*=1, 2, ……8 drawn from the Gaussian distribution with σ=0, 0.5, 1, 1.5, 2, 4, 6, 8 and 10.

Higher the time up to which the growth process is observed, greater will the separation between the Shannon entropy values. This can be further clearly observed with the help of the 3D contour plot of the evolution of the Shannon entropy values. The effect of noise fluctuations in the system can be further observed with the help of 2D contour plots. The contour plots clearly deciphers that there is a certain level of noise in the system after which no directionality can be observed and furthermore the transition from non-directional to directional fractals gets shifted with increment in the noise level in the system.

It is interesting to note that the fractal formation under noise fluctuations can be considered as a two state process, with two states being the s=0, where no noise is there and sd=10, where sufficient amount of noise is there which hinder the effect of noise reduction in the system. In simulation, with increase in the noise reduction parameter, one can finally achieve the directional fractals with high value of noise reduction parameter, even in the presence of very high amount of noise fluctuations. However, in experiments there is limit up to which the noise reduction can be achieved, indicating that directionality is not expected after certain level of noise in the system.

## 2. Morphological evolution of fractals

## a. Simulative evolution of the fractals growth

The evolution of the fractals, both in simulated (Fig. 13) and experimental evolution (Fig. 14) of the fractals was captured by calculating the Shannon entropy of the snapshots of the in between images of the video. The principle of distinguishing the morphology through Shannon entropy calculation is the change in the texture of the image at each growth step. At the early stage of evolution, the Shannon entropy values are same and hence one cannot distinguish the fate of the morphology at the end of the growth process. The bifurcation starts at a particular frame from where the change in the texture of the image can be captured by the imaging tool and the Shannon entropy values starts coming different (Fig. 15).

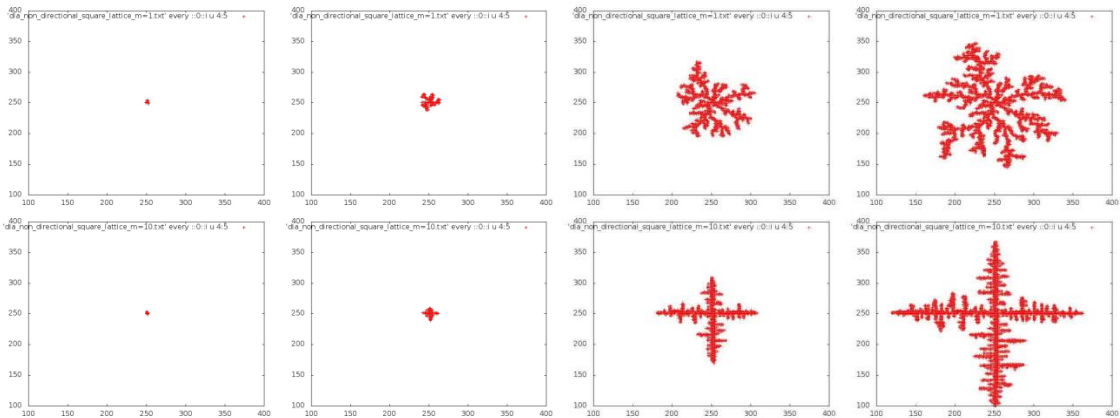

**Figure13 (colour online).** The figure represents the simulated fractal evolution for the isotropic and anisotropic fractal growth process. The isotropic growth process corresponds to m=1 and the anisotropic growth process corresponds to m=8.

**b. Experimental evolution of the fractals growth**

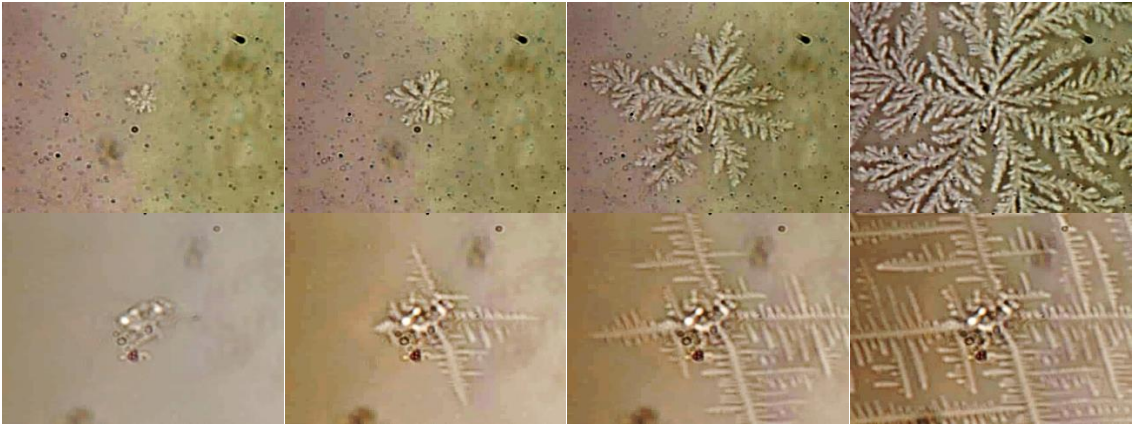

**Figure14 (colour online).** The figure represents the experimental fractal evolution for the isotropic and anisotropic fractal growth process.

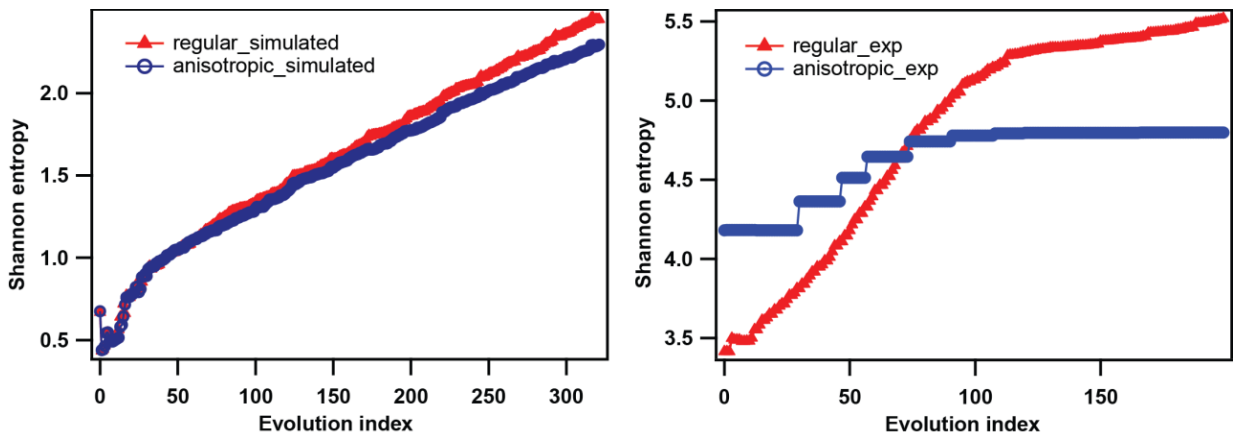

**Figure15 (colour online).** Shannon entropy calculation for the simulated and the experimental fractal evolution in the isotropic (red) and the anisotropic (blue) growth process.

Fig. 16 represents the calculation of the interaction energy of the formed fractals for the simulated fractals with different values of noise reduction parameter. The decrement in the interaction energy indicates towards the increasing stability of the formed fractals with the increase in the anisotropy in the morphology of the formed fractals. The Shannon entropy was observed to decrease in the similar fashion which confirms that Shannon entropy can be used for qualitatively inferring about the change in the interaction energy during the growth process. Moreover, the Shannon entropy was shown to capture the transition from the non-directional to directional fractals. Another important parameter for characterization of fractals is the fractal dimension. The fractal dimension also decreases with the increase in the noise reduction parameter. However, through fractal dimension no inference of the transition between the morphologies can be made, but the same can be very clearly inferred from the values of Shannon entropy (see Fig 16(b)).

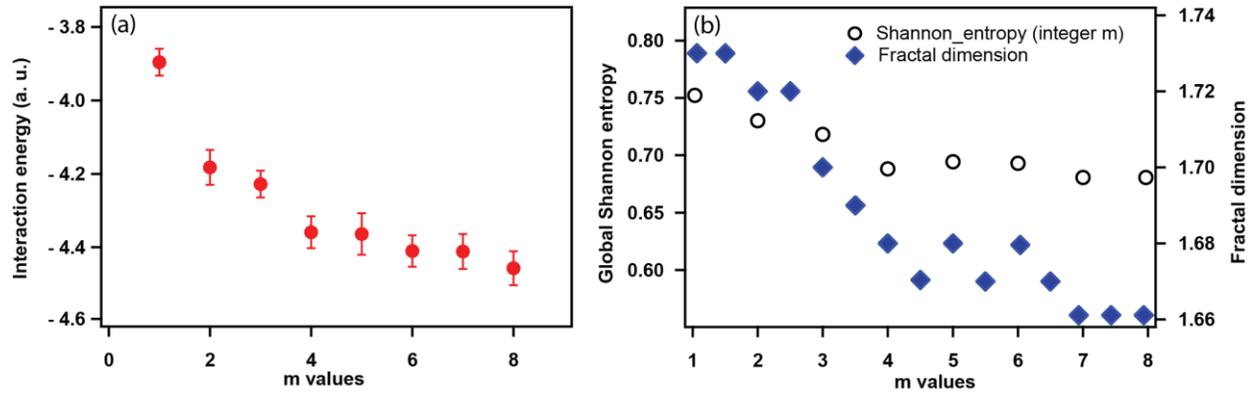

**Figure16 (colour online).** (a) The energy minimization curve for the fractals obtained with different noise reduction parameter on the square lattice. With increase in the noise reduction parameter, the interaction energy of the formed fractal is observed to be decreasing. (b) Shannon entropy calculation for the simulated fractals with noise fluctuations and the corresponding fractal dimension.

We calculated the interaction energy of the formed fractals with the eq. $V = -\frac{2}{3}\frac{\mu_A^2\mu_B^2}{(4\pi\epsilon)^2 r^6}\frac{1}{k_B T}$ (Fig. 16(a)). The decrease in the interaction energy with the increase in the $m$ values indicates towards the increase in the stability of the fractal aggregates with the increase in the directionality of the structure. Thermodynamically, the Gibbs' free energy ($G$) in terms of enthalpy ($H$), temperature ($T$) and entropy ($S$) of the system is given by $G=H-TS$ and under hard sphere interaction approximation, the contribution of enthalpy is negligible. The more negative is the Gibbs' free energy more favourable is the interaction and hence more negative is the interaction energy. In Fig. 16(a), it is clear that with increasing $m$ values, the interaction energy is becoming more negative and hence given the temperature of the system is constant, the entropy of the system will decrease and hence the system will become more ordered.

## 3. Analysis of cluster-cluster aggregation

We present here a model which accounts for the readjustment in the form of reorientations that the smaller cluster takes around the larger cluster in the cluster-cluster aggregation process. The reorientations in our model is mainly due to polarization effects which is explained by tip to tip cluster aggregation model in past.

### a. Theoretical background

While forming a fractal structure, the effective forces acting between the particles at the microscopic scale govern the pre-nucleation, nucleation and determine the morphology of structure. In past, DLVO potential (includes van der Waal and electrostatic double layer interaction) between the particles has been employed to explain the short-range structure of particle aggregation[39, 40]. A generalized theory of van der Waals was presented by McLachlan in 1964 [41] where the van der Waals free energy of two molecules 1 and 2 in a medium is given by the series,

$$w(r) = -\frac{6kT}{(4\pi\varepsilon_0)^2 r^6} \sum_{n=0,1,2\ldots}^{\infty} \frac{\alpha_1(iv_n)\alpha_2(iv_n)}{\varepsilon^2(iv_n)} \tag{1}$$

Where $\alpha_1(iv_n)$ and $\alpha_2(iv_n)$ are the polarizabilities of molecules 1 and 2, and $\varepsilon(iv_n)$ is the dielectric permittivity of the medium, at imaginary frequencies $iv_n$, where

$$v_n = (2\pi kT/h)n \approx 4 \times 10^{13} \, n \, s^{-1} \quad \text{at 300 K} \tag{2}$$

It should be noted that the zero frequency term, $n = 0$, is multiplied by ½. In case of molecules having single absorption frequency, the ionization frequency, $v_I$, its electronic polarizability at the real frequency is given by the damped simple harmonic oscillator model with $\kappa$ as the damping coefficient,

$$\alpha(v) = \alpha_0/[1 + i\kappa(v/v_I) - (v/v_I)^2] \tag{3}$$

In case of AuNP, the SPR shows the absorption maxima at 520 nm (see S2) which falls in the visible range. For visible range frequencies$(v_{vis}/v_I)^2 \ll 1$, $\alpha \approx \alpha_0$. In our framework, we assume that the larger clusters of AuNP, which becomes stationary, develop a permanent dipole moment $\vec{u}$ along the major axis of the cluster. The smaller clusters of AuNP which can move in space behave as polarizable molecules with a polarizability $\alpha_0$, a contribution from electronic polarizability only. The total polarizability of the large cluster is the sum of electronic and dipole polarizability and is given by

$$\alpha(iv_n) = \alpha_0/[1 - \kappa(v_n/v_l) + (v_n/v_l)^2] + u^2/3kT(1 + v_n/v_{rot}) \qquad (4)$$

where $v_{rot}$ is the average rotational relaxation frequency of the molecule. At zero frequency, the polarizability for larger cluster becomes $\alpha(0) = \alpha_0 + u^2/3kT$ (Debye-Langevin equation) and for the smaller cluster $\alpha(0) = \alpha_0$. The first term in the McLachlan's expression (eq. (1)) gives the zero frequency contribution to the free energy between larger and smaller AuNP cluster (neglecting higher order term),

$$w(r)_{v_n=0} = -\frac{3kT}{(4\pi\varepsilon_0)^2 r^6}\left[\frac{u^2}{3kT} + \alpha_0\right]\alpha_0 \approx -\frac{u^2\alpha_0}{(4\pi\varepsilon_0)^2 r^6} \qquad (5)$$

In the above expression, $k$ is the Boltzmann constant, $T$ is the environmental temperature, $\varepsilon_0$ is the permittivity of the free space, $r$ is the separation between centres of two clusters, $u$ is the dipole moment of the larger cluster and $\alpha_0$ is the polarizability of the smaller cluster. The above expression is obtained by averaging over orientations of the smaller clusters with Boltzmann distribution. The equation essentially represents the interaction between dipole and induced dipole and can be rewritten in an expression where the average over orientations has not been taken,

$$w(r) = -\frac{u^2\alpha}{2(4\pi\varepsilon_0)^2 r^6}(1 + 3cos^2\theta) \quad where \ 0 < \theta < \pi/2 \qquad (6)$$

The fractal aggregate formation is a complex process which starts with the formation of smaller clusters which finally grow into a larger fractal aggregate[42]. When a small cluster of AuNP approaches towards a bigger cluster of AuNPs, the interaction energy between these two for the zero frequency can be estimated using the equation

$$w = -\frac{u^2\alpha(4\,p_{orient})}{2(4\pi\varepsilon_0)^2 r^6} \qquad (7)$$

where $r$ is the separation between the centres of two clusters. In order to quantify the orientation effect, we define the orientation sticking probability (OSP) as the ratio of interaction potentials having any other orientation to interaction potential corresponding to most energetically favourable orientation, which can be represented as,

$$p_{orient} = \frac{(1+3\cos^2\theta)}{4} = \frac{2}{N_{reorient}} \; where \; 0 < \theta < \pi/2 \qquad (8)$$

where $N_{reorient}$ is the parameter related to the number of reorientations the smaller clusters take before getting attached to the bigger growing structure. Note that since the interaction equation has been written for zero frequency, for each value of $p_{orient}$, we have two values of $N_{reorient}$, and hence a multiplication factor of 2. This is because of the $\cos^2\theta$ term which makes the interaction energy same for the angles $\theta$ and $90° + \theta$, but in terms of parameter $N_{reorient}$, these two angles are visualized as two different orientations. Physically, the $p_{orient}$ represents the probability that the smaller cluster of AuNP will attach to the larger cluster of AuNP, given the orientation with which the smaller cluster is approaching the larger cluster is known. The value of OSP varies from 0.25 to 1.0 which shows that even if a smaller cluster is approaching with a least favourable orientation it has 25% probability of getting attached to the larger structure. This OSP evolves to 100% with the evolution of orientation from least energetically favourable to most energetically favourable orientation.

To observe the effect of OSP in simulation, we have modified DLA model by introducing an additional parameter $N_{reorient}$ in the well-known DLA model. We call this model as "orientation induced DLA model". In our simulation of cluster-cluster aggregation, the index of the neighbouring unoccupied lattice point has to be increased to $N_{reorient}$ rather than 1 for the particle to get attached to the growing structure. We call each instance of increase in the index value as one reorientation around that lattice point. It is important to note that the probability of the smaller cluster to get attached to the bigger cluster increases with each instance of reorientation until the index of the lattice point attains the value of $N_{reorient}$. The fact that parameter $N_{reorient}$ is this increased index value which the neighbouring unoccupied lattice point must attain before the particle can attach to the growing fractal structure relates it to OSP. Lower the value of OSP higher is the value of $N_{reorient}$, indicating more number of reorientations is required before attachment.

Also, it is important to note that there is no loss of generality in our approach compared to the observation through experiments. In the experiments, the smaller cluster approaches the larger structure and reorients itself to the orientation which corresponds to minimum energy configuration. The fact that particle in simulation has to approach $N_{reorient}$ times before getting attached to the larger structure attributes for the reorientations happening in the experiment.

For the case $N_{reorient}$=1, no reorientations are required and the smaller cluster will attach to the larger cluster in the first instance itself, hence represents the standard DLA model. For the orientation $\theta$= 0, $p_{orient} = 1$ and $N_{reorient}$=2. This shows that if the smaller cluster is approaching with the orientation which favors the minimum energy configuration then it has 100% probability of getting attached to the bigger cluster, thus $(N_{reorient} - 1) = 1$, the minimal number of reorientations required, hence $N_{reorient}$=2 and the smaller cluster can attach to the bigger cluster in the first instance of reorientation itself. Similarly for the approaching orientation $\theta$= 90°, $p_{orient} = 1/4$, and $N_{reorient}$=8. This shows that if the smaller cluster is approaching with the orientation which least favours the minimum energy configuration then it has 25% probability of getting attached to the bigger cluster, hence $(N_{reorient} - 1) = 7$ reorientations are required, hence $N_{reorient}$=8, and the smaller cluster can attach to the bigger cluster only after seven instances of reorientations. The parameter $N_{reorient}$, here used in cluster-cluster aggregation, is very similar to the noise reduction parameter $m$ used in particle-cluster aggregation for enhancing the inherent anisotropy of the lattice present in the growth process in on-lattice DLA process. This is evident from Fig. 2, where the change in the $m$ corresponds to the number of reorientations required before attachment leading to a clear transition in the morphology of the fractal aggregates.

## b. Experimental evidence

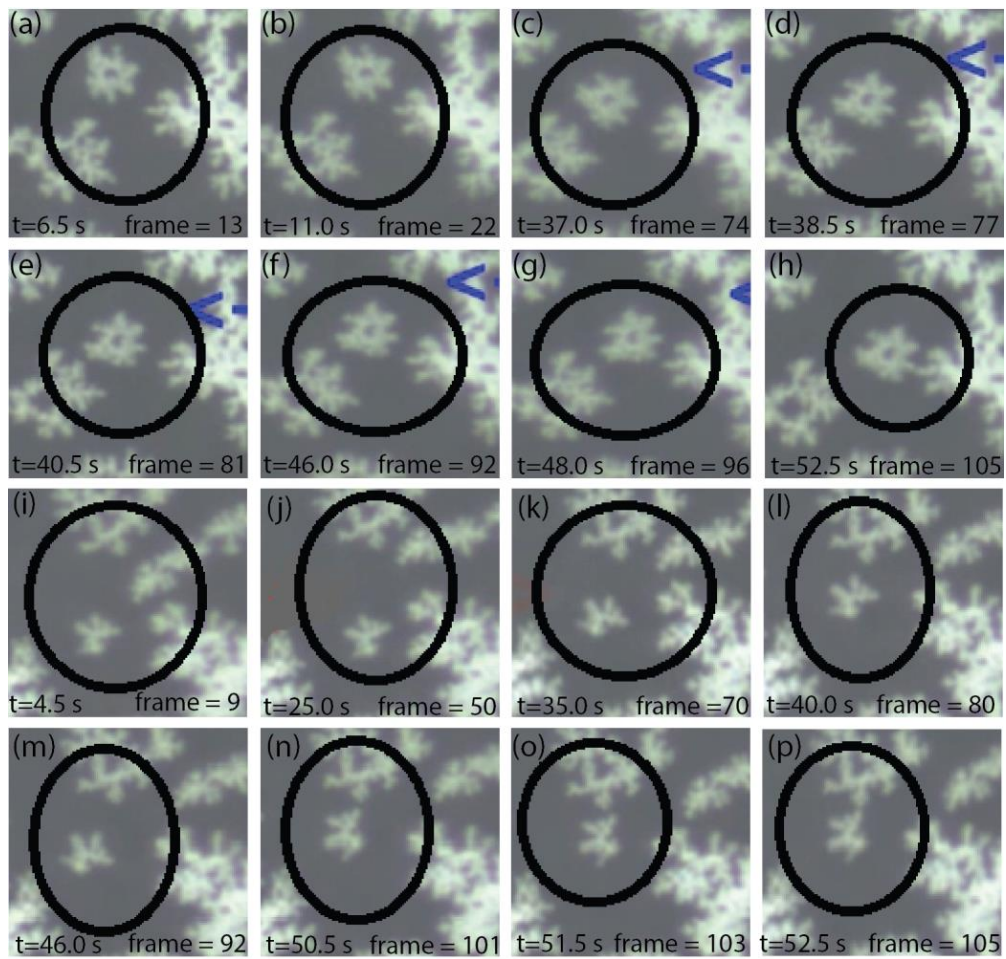

**Figure17 (colour online).** The figure shows the reorientations of two smaller clusters in the vicinity of the larger cluster. The first eight panels show the reorientation that the first (upper) cluster takes before getting attached with the bigger cluster. The next eight panels show the reorientation that the second (lower) cluster takes before getting attached with the bigger cluster. Each panel is the snapshot of the video (see supplementary video) containing the total reorientation effect. The black ellipse indicates the regions where the smaller cluster can possibly attach.

We observed the orientation effect experimentally in AuNP aggregates where smaller clusters were observed to reorient in the vicinity of the larger AuNP cluster. Figure 17 represents the eight such orientations for two smaller AuNP clusters. In Fig. 17 (a), three bigger clusters in the vicinity of the smaller cluster are shown where the smaller cluster can possibly attach. Through Fig. 17 (b)-(g), the smaller cluster reorients in the search of the best orientation to attach with and finally in Fig. 17 (h), the smaller cluster attaches to one of the bigger clusters with the orientation which corresponds to a minimum energy configuration. Similarly, in Fig. 17 (i)-(p), the same reorientation effect has been observed with another smaller cluster of AuNP. Fig. 17(i), as in Fig. 17 (a), shows the possible bigger clusters to which the smaller cluster can possibly attach. Fig. 17 (j)-(o) shows the reorientations the second cluster performed before finding the best orientation to attach with, which corresponds to a minimum energy configuration (Fig. 2 (p)).

The orientation with which the smaller cluster approaches the bigger cluster has lower $p_{orient}$. Hence, it performs reorientations for finding the best orientation with which it can attach so that the system attains the minimum energy. This reorientation effect in the experiment is related to the $N_{reorient}$ values in the simulation. This means the higher number of reorientations required in experiments corresponds to the higher value of $N_{reorient}$ in the simulation. The reorientation effects are related to orientational entropy associated with the system.

The order and entropy in a system are interrelated. This has been shown in previous work where the role of orientational entropy was seen by performing simulations with rods with different orientations[43]. In the first case, the aggregation was seen with rods with random (isotropic) orientations. However, when the rods were aligned (anisotropic), the probabilities of finding a configuration '$i$' having particular orientation of rods, $P_i$, were not the same and resulting entropy was found to be smaller than the isotropic case. In our case, the smaller AuNPs can take $N_{reorient}$, possible orientations which can be computed by the Eq. (8) and the entropy for the isotropic case and the anisotropic case can be mathematically given by $S_{isotropic} = -k_B \ln(N_{reorient})$ and $S_{aligned} = -k_B \int_1^{N_{reorient}} P \ln(P) dN_{reorient} \leq S_{isotropic}$ respectively where $N_{reorient}$ is the total number of orientations that smaller clusters of AuNPs can take and $P_i$ are probabilities of the smaller clusters to be in orientation '$i$' between 1 and $N_{reorient}$. The anisotropy in the cluster structure originates from anisotropy in the individual AuNPs. As is evident from the TEM images (see Fig 1(a)), at the structural level, the individual AuNPs are asymmetric and thus forms structurally asymmetric clusters. The orderliness of the formed fractal aggregates from these clusters depends on the orientations at which these individual clusters approach the growing aggregate.

**Table1.** The Shannon entropy calculated for simulated images (Fig 2) and the experimental images (Fig 17). The image index values are according to the image index in Figure2 in the main text.

| | Simulated images | | | | | Experimental images | | |
|---|---|---|---|---|---|---|---|---|
| m | FD | Shannon entropy | m | Shannon entropy | Image index | Upper cluster | Image index | Lower cluster |
| 1 | 1.73 | 0.7522 | 1.5 | 0.6790 | (a) | 5.5285 | (i) | 5.6681 |
| 2 | 1.72 | 0.7302 | 2.5 | 0.6602 | (b) | 5.5954 | (j) | 5.4785 |
| 3 | 1.70 | 0.7182 | 3.5 | 0.6312 | (c) | 5.7106 | (k) | 5.7849 |
| 4 | 1.68 | 0.6882 | 4.5 | 0.5922 | (d) | 5.6859 | (l) | 5.8037 |
| 5 | 1.68 | 0.6944 | 5.5 | 0.5715 | (e) | 5.6444 | (m) | 5.6762 |
| 6 | 1.68 | 0.6943 | 6.5 | 0.6028 | (f) | 5.6325 | (n) | 5.6740 |
| 7 | 1.66 | 0.6929 | 7.5 | 0.5569 | (g) | 5.5954 | (o) | 5.6183 |
| 8 | 1.66 | 0.6807 | | | (h) | 5.5673 | (p) | 5.7747 |

We applied the image processing tools to partly explain the entropy variation with different cluster reorientations. We verified the feasibility of application of Shannon entropy concept for estimation of entropy in our experimental images (Figure 16(b)). We first calculated the Shannon entropy for simulated fractal structures and observed that the entropy values followed the same trend as that of the fractal dimension (see Fig 16(a)), hence indicating that the Shannon entropy parameter can be used for qualitatively estimating the role of reorientations in changing the entropy in case of experimental images.  In case of simulated fractals (Fig. 2), the Shannon entropy decreased with increasing anisotropy in the fractal structure, 0.7250 for $m$=1 and 0.6061 for $m$=8. In case of experimental images, the Shannon entropy partly captured the reorientation effects and showed the increase in entropy for the final case where the smaller cluster got attached to the larger cluster (lower cluster, first image- 5.6681, last image- 5.7747 and upper cluster, first image- 5.5285, last image- 5.5673) (see Table 1). The change in entropy for both simulated fractal images and the experimental images is shown in the inset of Fig. 18.

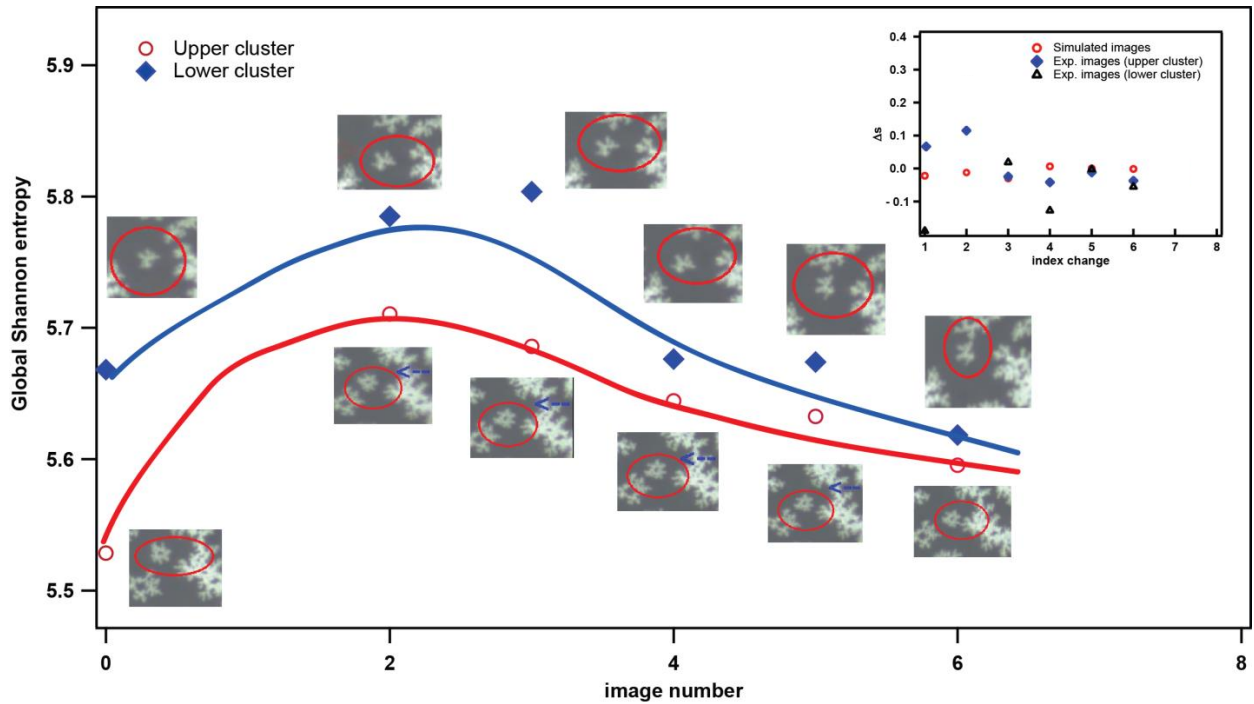

**Figure18 (colour online).** The figure shows the global Shannon entropy experimental images (both upper and lower cluster). The index number on the x-axis corresponds to the images index given in the figure (2). Along with the data points shown are the filtered images of the experimental images shown in figure (2) obtained with "entropyfilt" Matlab routine. The inset in the figure shows the global Shannon entropy change for simulated images, experimental images (both upper and lower cluster). The index change on the x-axis means the transition between subsequent $m$ values (for simulated images) and transition between subsequent images (for experimental images). The red and the blue curves highlight the direction of entropy change with the reorientations.

As shown in Fig 18, the Shannon entropy partly captured the entropy changes in the system due to reorientations. For both the clusters, there was increase in the Shannon entropy when the smaller clusters moved away from the larger cluster (transition from Fig. 17(b)-(c) for upper cluster and Fig. 17(j)-(k) for lower cluster). This indicates that the configuration where the smaller cluster is away from the larger cluster is energetically less favourable compared to the configuration where the smaller cluster moves towards the larger cluster (transition from Fig. 17(e)-(f) for upper cluster and Fig. 17(m)-(n) for lower cluster) in the process to attach it. The change in the entropy due to reorientations is captured in transitions from Fig. 17(c)-(d) for upper cluster and Fig. 17(k)-(l) for lower cluster where the increase (decrease) in the entropy for lower (upper) cluster can be seen as the effect of the change in the orientation of the smaller cluster around the larger cluster.

**Conclusions**

In the first part of this work, the particle aspect ratio based noise fluctuations in the system are addressed with Shannon entropy. It has been shown that noise fluctuations delayed the transition from where the transition from isotropic to anisotropic fractals will start. Also, the analysis indicates that in real systems no directionality in the morphologies is expected after certain value of noise fluctuations in the system. The results indicate the process of fractals formation to be a two state process in the presence of noise fluctuations.

As the second objective of this work, the feasibility of application of Shannon entropy for capturing the morphological evolution is checked and the imaging tool was found to be convincingly applicable in capturing the emergence of the anisotropy in the growth process. The analysis shows that there is cut-off value of the difference in the texture of the image after which the Shannon entropy can appreciably distinguish between the isotropic and the anisotropic growth processes. This cut-off is reflected as the bifurcation point in the calculated Shannon entropy values for isotropic and the anisotropic morphologies.

The imaging tool is shown to be capable of qualitatively accounting the increase in the stability of the formed fractals with increase in the noise reduction parameter. Moreover, the fractal dimension cannot capture the transition of the morphologies, but Shannon entropy has been shown to indicate the value of the noise reduction parameter after which the transition from isotropic to anisotropic fractal growth starts.

Finally in this work, the 'orientation induced DLA model' for cluster-cluster aggregation is applied to explain the role of orientations in providing different structural features to growing colloidal aggregates. The OSP is quantitatively included in simulation by including a parameter $N_{reorient}$ in DLA model, very similar to the noise reduction parameter used in particle-cluster DLA model, which takes care of the reorientations happening in the actual system. With the increase in the value of $N_{reorient}$, the directionality in the grown fractal structures is observed. This indicates that higher the value of $N_{reorient}$ greater is the number of reorientations required for the smaller cluster to get attached to the growing structure to attain the minimum energy configuration.

**Acknowledgement**

AS acknowledges the support of institute fellowship and ANG acknowledges the support of ISIRD grant from the Indian Institute of Technology Kharagpur, Kharagpur, India.